\documentclass[12pt]{iopart}

\usepackage{graphicx}
\def\prl{Phys. Rev. Lett.}
\def\prd{Phys. Rev. D}
\def\cqg{Class. Quantum Grav.}
\def\lrr{Living Rev. Relat.}

\begin{document}
   
\title{Puncture black hole initial data in the conformal thin-sandwich formalism}

\author{Thomas W. Baumgarte\footnote{Also at Department of Physics, University of Illinois at
Urbana-Champaign, Urbana, IL 61801, USA}}
\address{Department of Physics and Astronomy, Bowdoin College,
  Brunswick, ME 04011, USA}

\ead{tbaumgar@bowdoin.edu}
   
\begin{abstract}
We revisit the construction of puncture black hole initial data in the conformal thin-sandwich decomposition of Einstein's constraint equations.  It has been shown previously that this approach cannot yield quasiequilibrium wormhole data, which connect two asymptotically flat spatial infinities.   This argument does not apply to trumpet data, which connect the spatial infinity in one universe with the future timelike infinity of another.   As a numerical demonstration we present results for a single boosted trumpet-puncture black holes, constructed in the original version of the conformal thin-sandwich formalism.  
\end{abstract}



\section{Introduction}

Numerical simulations of black hole spacetimes have experienced a dramatic breakthrough (see \cite{Pre05b,BakCCKM06a,CamLMZ06} as well as numerous later publications).  Many of these simulations now adopt some variation of the Baumgarte-Shapiro-Shibata-Nakamura formulation \cite{ShiN95,BauS98} together with the moving-puncture \cite{BakCCKM06a,CamLMZ06} method to handle the black hole singularities.  

Typically, moving-puncture simulations adopt initial data that are constructed using the {\em puncture} method \cite{BraB97}.  As we explain in more detail below, the central idea of the puncture method is to write the unknown functions as sums of analytically known background terms that capture the singularities, and correction terms that are unknown but regular.  If all goes well, the equations for the correction terms can then be solved everywhere, without any need for excision or other means of dealing with the black hole singularity.  Most applications of this method have adopted Schwarzschild data on a slice of constant Schwarzschild time as the background solution.  In the Penrose diagrams of Fig.~\ref{Fig1} below, these slices would be represented by lines connecting spatial infinity in one universe with spatial infinity of another universe.  The resulting initial data represent {\em wormhole} data.  Most applications to date have applied the puncture method in the context of the {\em conformal transverse-traceless} (CTT) decomposition of Einstein's constraint equations \cite{Yor79,Coo00,Alc08,BauS10}.

The {\em moving-puncture} method is used in dynamical simulations, and is based on a set of empirically found coordinate conditions, namely the ``1+log" slicing condition for the lapse \cite{BonMSS95} and a ``$\bar \Gamma$-freezing" gauge condition for the shift \cite{AlcBDKPST03}.  As demonstrated by \cite{HanHPBO06,HanHOBGS06,Bro08,HanHOBO08,DenWBB10}, dynamical simulations of a Schwarzschild spacetime using these coordinate conditions settle down to a spatial slice that terminates at a non-zero areal radius.  An embedding diagram of such a slice, which suggests the name {\em trumpet} data, is shown in Fig.~2 of \cite{HanHOBO08} (see also \cite{BauN07,Bru09} for an analysis of these slices).  We also include such a slice in the lower panel of Fig.~\ref{Fig1}, which demonstrates that trumpet slices connect spatial infinity in one universe with future timelike infinity of another universe.  {\em Trumpet-puncture} initial data, that adopt the puncture method in the CTT formalism and produce black holes in a trumpet geometry, have recently been constructed in \cite{ImmB09,HanHO09}.

While most initial data that are currently being used in moving-puncture simulations have adopted the CTT formalism, the {\em conformal thin-sandwich} (CTS) formalism \cite{Yor99,Coo00,PfeY03,Alc08,BauS10} offers an attractive alternative.  Several time derivatives can be set to zero in the CTS formalism, suggesting that the resulting data may represent quasiequilibrium initial data more faithfully than CTT data.   Moreover, the CTS formalism yields, as part of the solution, the coordinate system in which these time derivatives vanish.

Given these considerations it is of interest, at least as a matter of principle, to construct puncture initial data in the CTS formalism.  In \cite{HanECB03}, Hannam, Evans, Cook \& Baumgarte asked ``Can a combination of the conformal thin-sandwich and puncture methods yield binary black hole solutions in quasiequilibrium?" and concluded that this is impossible, at least for wormhole data that connect two spatial infinities.   Here we present numerical examples for boosted black holes to demonstrate that it is possible to construct quasiequilibrium trumpet-puncture initial data that connect spatial infinity in one universe with future timelike infinity of another, at least in the original version of the CTS formalism.  The extended version of the CTS formalism, which has been more widely used in numerical relativity applications, seems to introduce new complications in this approach.

This paper is organized as follows.   In Section \ref{sec:bas_equ} we discuss decompositions of 
Einstein's equations and introduce the equations of the CTS decomposition.  In Section \ref{sec:puncture} we review the puncture method, go over the arguments of \cite{HanECB03} to explain why it cannot be adopted within the CTS formalism to construct quasiequilibrium wormhole data, and argue that the situation is different for trumpet data.  We present numerical results for boosted black holes constructed in the original CTS decomposition in Section \ref{sec:num_res}.  
In Section \ref{sec:discussion} we summarize our results, and briefly discuss this approach in the context of the extended CTS decomposition, where new difficulties seem to arise.  We also include a brief Appendix that reviews the Schwarzschild solution in a maximally sliced trumpet geometry.

\section{Basic equations}
\label{sec:bas_equ}


In a 3+1 decomposition of general relativity (see, e.g., \cite{ArnDM62,Yor79,BauS10}), the spacetime $M$ is foliated by a family of spatial hypersurfaces $\Sigma$.  The ten Einstein equations for the spacetime metric $g_{ab}$ are projected into the spatial slices, which results in a set of constraint equations and a set of evolution equations.  The constraint equations constrain the gravitational fields on each spatial slice, while the evolution equations govern their evolution from one slice to the next.  Constructing initial data for Einstein's equations entails finding solutions to the constraint equations.

It is convenient to write the spacetime metric as
\begin{equation}
ds^2 = - \alpha^2 dt^2 + \gamma_{ij}(dx^i + \beta^i dt) (dx^j + \beta^j dt),
\end{equation}
where $\alpha$ is the lapse, $\beta^i$ is the shift vector, and the spatial metric $\gamma_{ij}$ is the projection of the spacetime metric $g_{ab}$ into the spatial slices $\Sigma$.  We also define the extrinsic curvature as
\begin{equation} \label{K_def}
K_{ij} \equiv - \frac{1}{2\alpha} \left( \partial_t \gamma_{ij} - D_i \beta_j - D_j \beta_i \right),
\end{equation}
where the operator $D_i$ denotes a covariant derivative with respect to $\gamma_{ij}$.  In vacuum, Einstein's equations are then equivalent to the two constraint equations
\begin{equation} \label{ham_const_1}
R + K^2 - K_{ij} K^{ij} = 0
\end{equation}
(the Hamiltonian constraint) and 
\begin{equation} \label{mom_const_1}
D_j (K^{ij} - \gamma^{ij} K) = 0
\end{equation}
(the momentum constraint), and the evolution equation
\begin{eqnarray} \label{K_dot}
\partial_t K_{ij} & =  & \alpha (R_{ij} - 2 K_{ik} K^k{}_j + K K_{ij}) - D_i D_j \alpha \nonumber \\
& & 	+ \beta^k \partial_k K_{ij} + K_{ik} \partial_j \beta^k + K_{kj} \partial_i \beta^k.
\label{Kij_dot}
\end{eqnarray}
Here $K \equiv \gamma^{ij} K_{ij}$ is the trace of the extrinsic curvature, also called the mean curvature, and $R_{ij}$ is the Ricci tensor associated with the spatial metric.  The above set of equations is often referred to as the Arnowitt-Deser-Misner, or ADM equations \cite{ArnDM62}.


The constraint equations (\ref{ham_const_1}) and (\ref{mom_const_1}) can be decomposed further with the help of a conformal transformation
\begin{equation}
\gamma_{ij} = \psi^4 \bar \gamma_{ij},
\end{equation}
where $\psi$ is a conformal factor and $\bar \gamma_{ij}$ a conformally related metric.  We also split the extrinsic curvature into its trace $K$ and trace-free part $A_{ij}$,
\begin{equation}
K_{ij} = A_{ij} + \frac{1}{3} \gamma_{ij} K,
\end{equation}
and conformally transform $A_{ij}$ according to
$
A_{ij} = \psi^{-2} \bar A_{ij}.
$
The Hamiltonian constraint (\ref{ham_const_1}) then becomes
\begin{equation} \label{ham_const_2}
\bar D^2 \psi - \frac{1}{8} \psi \bar R - \frac{1}{12} \psi^5 K^2 + \frac{1}{8} \psi^{-7} \bar A_{ij} \bar A^{ij} = 0,
\end{equation}
while the momentum constraint (\ref{mom_const_1}) is
\begin{equation} \label{mom_const_2}
\bar D_j \bar A^{ij} - \frac{2}{3} \psi^6 \bar \gamma^{ij} \bar D_j K = 0.
\end{equation}
Here $\bar D_i$,  $\bar D^2 \equiv \bar D^i \bar D_i$ and $\bar R$ are the covariant derivative, the covariant Laplace operator and the Ricci scalar associated with $\bar \gamma_{ij}$.  Different decompositions of the constraint equations proceed with different further decompositions of $\bar A_{ij}$.  


In the {\em conformal transverse-traceless} (or CTT) decomposition, $\bar A_{ij}$ is decomposed into a transverse and a longitudinal part.  The transverse part can then be chosen freely, while the longitudinal part has to satisfy the momentum constraint.  In addition to the transverse part of $\bar A_{ij}$, the mean curvature $K$ and the conformally related metric $\bar \gamma_{ij}$ are freely specifiable and have to be chosen before the equations can be solved.

The CTT decomposition as the very attractive feature that, for vacuum, maximal slicing ($K = 0$) and conformal flatness ($\bar \gamma_{ij} = \eta_{ij}$, where $\eta_{ij}$ is the flat metric in any coordinate system),  the momentum constraint decouples from the Hamiltonian constraint and can be solved analytically for both boosted and spinning black holes.  These Bowen-York solutions \cite{BowY80} can then be inserted into the Hamiltonian constraint, which is the only equation that needs to be solved numerically.



In the {\em conformal thin-sandwich} (or CTS) decomposition, we invoke eq.~(\ref{K_def}) to write $\bar A_{ij}$ as
\begin{equation} \label{A_1}
\bar A^{ij} = \frac{1}{2 \bar \alpha} \left((\bar L \beta)^{ij} - \bar u^{ij} \right),
\end{equation}
where we have introduced the ``densitized lapse" $\bar \alpha$ with the conformal rescaling $\alpha = \psi^6 \bar \alpha$, and have defined
\begin{equation}
\bar u_{ij} \equiv \partial_t \bar \gamma_{ij}.
\end{equation}
The longitudinal operator, or vector gradient $\bar L$ is
\begin{equation}
(\bar L \beta)^{ij} \equiv \bar D^i \beta^j + \bar D^j \beta^i - \frac{2}{3} \bar \gamma^{ij} \bar D_k \beta^k.
\end{equation}
Inserting (\ref{A_1}) into the momentum constraint (\ref{mom_const_2}) yields
\begin{equation} \label{mom_const_3}
(\bar \Delta_L \beta)^i - (\bar L \beta)^{ij} \bar D_j \ln \bar \alpha = 
\bar \alpha \bar D_j (\bar \alpha^{-1} \bar u^{ij}) + \frac{4}{3} \bar \alpha \psi^6 \bar D^i K,
\end{equation}
where 
\begin{equation}
(\bar \Delta_L \beta)^i \equiv \bar D_j (\bar L\beta)^{ij}
= \bar D^2 \beta^i + \frac{1}{3} \bar D^i (\bar D_j \beta^j) + \bar R^i{}_j \beta^j
\end{equation}
is the vector Laplacian, and $\bar R^i{}_j$ the Ricci tensor associated with $\bar \gamma_{ij}$.

The Hamiltonian constraint (\ref{ham_const_2}) and the momentum constraint (\ref{mom_const_3}) form the basic equations of the CTS decomposition in its {\em original} form (\cite{Yor99}, see also Box 3.2 in \cite{BauS10}).  The freely specifiable variables in this decomposition are the conformally related spatial metric $\bar \gamma_{ij}$, its time derivative $\bar u_{ij}$, the mean curvature $K$, and the densitized lapse $\bar \alpha$.  Choices for these functions can be inserted into eqs.~(\ref{ham_const_2}) and (\ref{mom_const_3}), which can then be solved for the conformal factor $\psi$ and the shift vector $\beta^i$.  Given these solutions, the physical solutions for $\gamma_{ij}$ and $K_{ij}$ can be constructed by reversing the steps above.   For $\bar \alpha = 1/2$ and $\bar u_{ij} = 0$, the equations of the original CTS formalism reduce to those of the CTT formalism (given suitable choices of the free variables in that formalism); in particular, the Bowen-York solutions are solutions to the original CTS system under these conditions (see \cite{HanHBGS07}). 


In the {\em extended} CTS decomposition (see, e.g., \cite{PfeY03}), eqs.~(\ref{ham_const_2}) and (\ref{mom_const_3}) are supplemented with the trace of the evolution equation (\ref{Kij_dot}).  Combining this trace with the Hamiltonian constraint (\ref{ham_const_2}) we obtain an equation
\begin{equation} 
\bar D^2 \phi = \phi \left( \frac{7}{8} \psi^{-8} \bar A_{ij} \bar A^{ij} +
\frac{5}{12} \psi^4 K^2 + \frac{1}{8} \bar R \right) - \psi^5 \partial_t K + \psi^5 \beta^i \bar D_i K.  \label{lapse_1}
\end{equation}
for the product of the lapse $\alpha$ and the conformal factor $\psi$,  $\phi \equiv \alpha \psi = \bar \alpha \psi^7.$ Instead of the conformally rescaled lapse $\bar \alpha$, the time derivative of the mean curvature $\partial_t K$ (together with $K$, $\bar \gamma_{ij}$ and $\bar u_{ij}$) now acts as freely specifiable variable (see also Box 3.3 in \cite{BauS10}).  

\section{The puncture method}
\label{sec:puncture}

Solving any of the above equations for black holes requires handling the singularities in the black hole interior.  

One approach is to excise the black hole interior with the help of suitable boundary conditions on the black hole horizon (see, e.g., \cite{CooY90,Coo94} in the context of the CTT decomposition, and \cite{GouGB02,GraGB02,CooP04,CauCGP06} for the CTS decomposition).   An attractive feature of the excision method is that the boundary conditions can be derived from geometrically motivated conditions, and can be used to force the black hole to be in equilibrium.  However, this approach can lead to complicated numerical grids, and can be tedious to implement numerically.  Moreover, this approach results in valid data only in the black hole exterior, while moving-puncture evolution codes need initial data everywhere (compare \cite{EtiFLSB07,BroSSTDHP07}).

The puncture method, on the other hand, is much easier to implement numerically.  The central idea of the puncture method is to write all functions as sums of a background term, describing a Schwarzschild black hole, and a correction term (see \cite{BraB97}, see also \cite{BeiO94,BeiO96}).  The background terms are given analytically and absorb the singularities, while, if all goes well, the correction terms remain regular everywhere and can be found numerically, without the need of excision, on $R^3$.  

\begin{figure}[t]
\begin{center}
\includegraphics[width=3in]{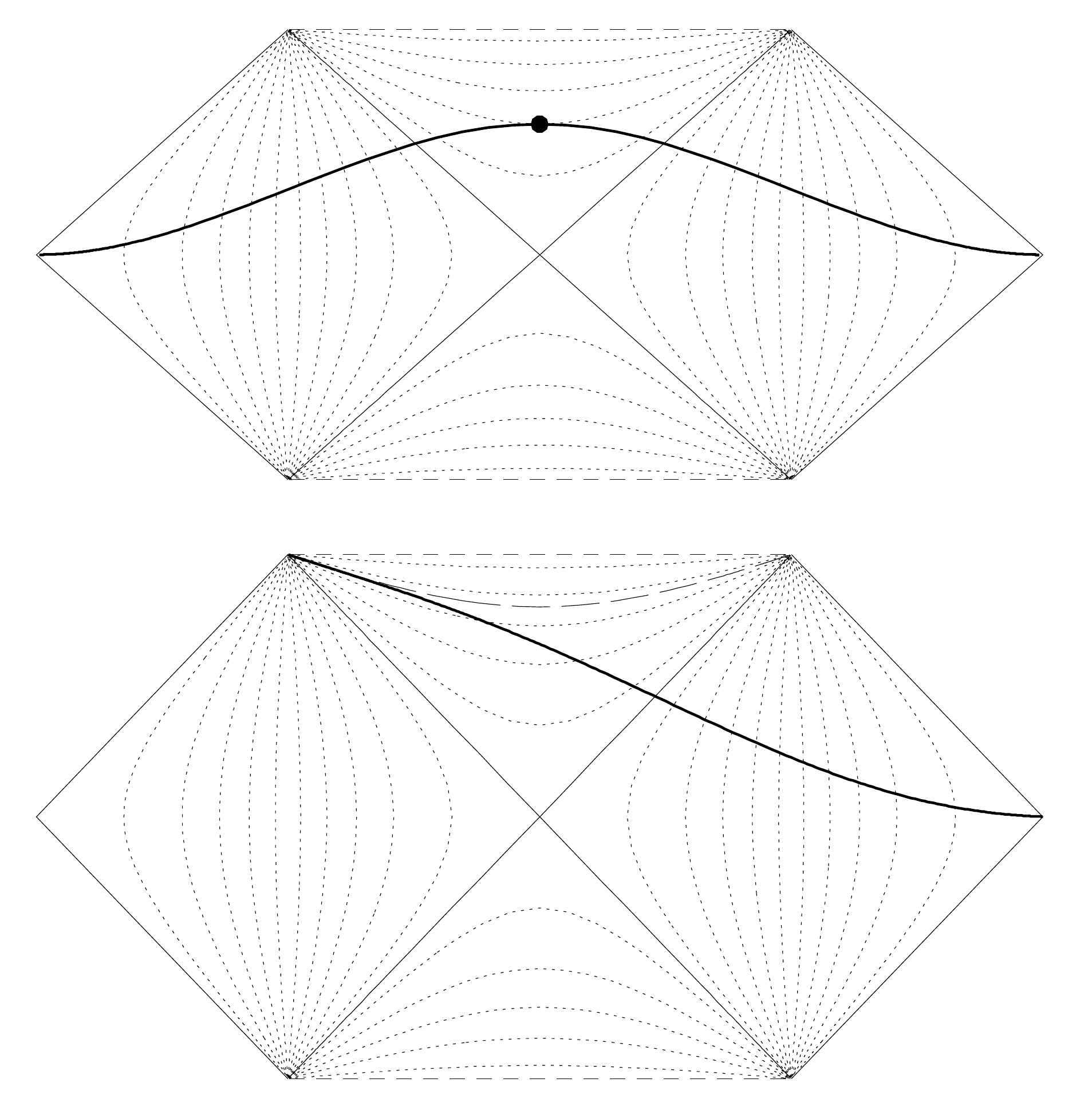}
\end{center}
\caption{Penrose diagrams for a Schwarzschild black hole.  The dotted lines are lines of constant areal radius $R$; the Killing vectors $\xi_{(t)}^a$ are tangent to these lines.  The thick solid line in the top panel represents a spatial ``wormhole" slice connecting the spatial infinities in two separate universes.  The thick solid line in the bottom panel represents a spatial ``trumpet" slice.  This slice starts at the asymptotically flat spatial infinity of one universe, enters the black hole region, and asymptotes to the future timelike infinity of the other universe without ever leaving the black hole region.  The long-dashed line in the lower panel marks the limiting surface at $R = 3M/2$.}
\label{Fig1}
\end{figure}

The Schwarzschild background terms can be given in a number of different coordinate systems.  Most applications to date have adopted the data on a slice of constant Schwarzschild time (e.g.~\cite{BraB97,Bau00}); these data connect spatial infinity in one universe with spatial infinity in another universe; the resulting initial data therefore represent {\em wormhole} data (see the top panel in Fig.~\ref{Fig1}).  

It is also possible to adopt maximally-sliced {\em trumpet} data as background data in the puncture method \cite{ImmB09,HanHO09}.  These data appear as the limiting case of the family of time-independent, maximal slices of the Schwarzschild spacetime (see, e.g., \cite{EstWCDST73,Rei73,BeiO98}, see also \ref{AppA}).  Trumpet data connect spatial infinity in one universe with future timelike infinity of another universe (see the bottom panel in Fig.~\ref{Fig1}; see also Fig.~4 in \cite{HanHOBO08}), and approach asymptotically a sphere of proper radius $R = 3M/2$.  When displayed in an embedding diagram, these slices resemble the shape of a trumpet (see Fig.~2 in \cite{HanHOBO08}), which motivates the name of the resulting geometry.  As we discussed in the introduction, these data have the appealing feature that they are close to the geometry realized in dynamical simulations in the moving-puncture formalism, and avoid a coordinate transition from a wormhole to a trumpet geometry.  It turns out that these data also have very interesting properties from the perspective of the CTS formalism.

To date, the puncture formalism has been adopted almost exclusively in the context of the CTT decomposition, where only the Hamiltonian constraint needs to the solved numerically (see \cite{HanC05,Han05,Lag04} for exceptions).  Given the success of the puncture method in the CTT decomposition, it is tempting to apply the same method in the CTS formalism.  In \cite{HanECB03}, the authors therefore asked ``Can a combination of the conformal thin-sandwich and puncture methods yield binary black hole solutions in quasiequilibrium?" and concluded that this is impossible, at least for wormhole data.  The argument goes as follows.
 
 Consider a Penrose diagram for a Schwarzschild black hole, as in Fig.~\ref{Fig1}.  The top panel in the figure includes a spatial wormhole slice, which connects the spatial infinities in two separate universes.  Any such slice must, at some point, be tangent to a surface of constant areal radius $R$, marked as the dotted lines in the diagram.  This point, marked by a dot in Fig.~\ref{Fig1}, defines the slice's minimal surface, or its ``throat".  Note that the Killing vector $\xi_{(t)}^a$, which is timelike in the exterior of the black hole but spacelike inside the horizon, is tangent to the surfaces of constant $R$.  For equilibrium data, we require that the time vector 
 \begin{equation} \label{t_Killing}
 t^a = \alpha n^a + \beta^a,
 \end{equation}
 where $n^a$ is the normal on the spatial slice, be aligned with the Killing vector $\xi_{(t)}^a$; this defines the ``Killing lapse" $\alpha_K$ and the ``Killing shift" $\beta_K^i$.  At the throat, $t^a$ must therefore be tangent to the slice, which implies that at this point the Killing lapse $\alpha_K$ must vanish (see, e.g., equation (\ref{wormhole_Killinglapse}) below).

Unfortunately, the CTS formalism requires dividing by the lapse in equation (\ref{A_1}), so that the equations cannot be solved without special treatment of the throat.  This defeats the purpose of the puncture method.  The authors of \cite{HanECB03} therefore concluded that is is impossible to construct quasiequilibrium, wormhole puncture initial data in the CTS decomposition.  One way to avoid these problems is to relax the assumption of quasiequilibrium, and make different choices for the lapse (see \cite{HanC05,Han05}; see also Section \ref{sec:cts_org_worm} below).  

Another way of avoiding these problems is to relax the assumption that the method be based on wormhole data that connect the two spatial infinities of two separate universes.   We now see that trumpet data offer a very attractive alternative in this context.  Since these slices do not connect two spatial infinities, but rather a spatial infinity in one universe with a future timelike infinity of another universe, these slices become tangent to the surfaces of constant $R$ only asymptotically (see also \cite{HanHPBO06}).  This is demonstrated in the lower panel of Fig.~\ref{Fig1}, which includes such a trumpet slice.  Accordingly, the Killing lapse vanishes only at the end of the slice, where singularities are absorbed in the ``puncture".  This observation suggests to try to construct trumpet-puncture initial data in the CTS formalism.  

\section{Numerical Results}
\label{sec:num_res}


In the original CTS decomposition 
we need to solve the Hamiltonian constraint (\ref{ham_const_2}) and the momentum constraint (\ref{mom_const_3}) together with expression (\ref{A_1}) for the extrinsic curvature.  Before proceeding we need to make choices for the freely specifiable variables; we will assume conformal flatness $\bar \gamma_{ij} = \eta_{ij}$ (which implies $\bar R_{ij} = 0$), maximal slicing $K = 0$, and, consistent with the notion of quasiequilibrium, we will assume $\bar u_{ij} = 0$.  We will postpone making choices for $\bar \alpha$ until we specialize to wormhole and puncture data in Sections \ref{sec:cts_org_worm} and \ref{sec:cts_org_trump} below.  With these choices, the Hamiltonian constraint (\ref{ham_const_2}) reduces to
\begin{equation} \label{ham_const_3}
\bar D^2 \psi = -  \frac{1}{8} \psi^{-7} \bar A_{ij} \bar A^{ij},
\end{equation}
while the momentum constraint (\ref{mom_const_3}) becomes 
\begin{equation} \label{mom_const_4}
(\bar \Delta_L \beta)^i - (\bar L \beta)^{ij} \bar D_j \ln \bar \alpha_0 = 0.
\end{equation}
The momentum constraint decouples from the Hamiltonian constraint and can be solved independently of a solution for the conformal factor $\psi$.  We decorate the densitized lapse $\bar \alpha_0$ with a subscript zero as a reminder that this quantity is held fixed in the original CTS formalism.  Given a solution for $\beta^i$ we can compute $\bar A^{ij}$ from (\ref{A_1}), which now becomes
\begin{equation} \label{A_1.5}
\bar A^{ij} = \frac{1}{2 \bar \alpha_0} (\bar L \beta)^{ij}.
\end{equation}

To apply the puncture method, we write the unknown variables as the sum of some background term and a correction term (but do not linearize in the correction term, so that the equations remain valid even in the non-linear regime).  In principle, any choice for the background terms is possible, but we will restrict our analysis to background choices that satisfy equations (\ref{ham_const_3}) through (\ref{A_1.5}) above.  Towards that end,  we pick a spatial slice of the Schwarzschild spacetime and adopt the data on this slice as background data.  We will assume that this slice is a member of the family of spherically symmetric, time-independent maximal slices of Schwarzschild (see \cite{EstWCDST73,Rei73,BeiO98}).  The geometry of this slice determines the spatial metric and the extrinsic curvature.  Adopting isotropic spatial coordinates we can identify the background conformal factor $\psi_0$ as well as the traceless, conformally related extrinsic curvature $\bar A^{ij}_0$.  Also associated with this slice is a Killing lapse $\alpha_K$ and a Killing shift $\beta_K^i$ which, in a dynamical evolution, would leave the background time-independent.  

We now write the conformal factor $\psi$ as 
\begin{equation} \label{psi_decomp}
\psi = \psi_0 + u.
\end{equation}
and the shift $\beta^i$ as
\begin{equation} \label{beta_decomp}
\beta^i = \beta^i_0 + b^i.
\end{equation}
Inserting (\ref{psi_decomp}) into the Hamiltonian constraint (\ref{ham_const_3}) yields
\begin{equation} \label{ham_const_4}
\bar D^2 u = - \frac{1}{8} (\psi_0 + u)^{-7} \bar A_{ij} \bar A^{ij} +
	\frac{1}{8} \psi_0^{-7} \bar A^0_{ij} \bar A^{ij}_0,
\end{equation}
where we have used the fact that the background conformal factor must satisfy the Hamiltonian constraint.  We also have
\begin{equation} \label{A_2}
\bar A^{ij} = \frac{1}{2 \bar \alpha_0} ( \bar L b)^{ij} + \frac{1}{2\bar \alpha_0} (\bar L \beta_0)^{ij}
= \frac{1}{2 \bar \alpha_0} ( \bar L b)^{ij} + \bar A_0^{ij},
\end{equation}
while inserting (\ref{beta_decomp}) into the momentum constraint (\ref{mom_const_4}) results in
\begin{equation} \label{mom_const_5}
(\bar \Delta_L b)^i - (\bar L b)^{ij} \bar D_j \ln \bar \alpha_0 = 0,
\end{equation}
since the background term $\beta_0^i$ satisfies the constraint (\ref{mom_const_4}) itself.  Equations (\ref{ham_const_4}) - (\ref{mom_const_5}) form the equations of the original CTS decomposition in the puncture approach.  Given choices for $\psi_0$, $\beta^i_0$ and $\bar \alpha_0$ we can now solve these equations for the corrections $u$ and $b^i$.

As demonstrated by Laguna \cite{Lag04}, solutions to the above equations for boosted (or spinning) black holes are geometrically equivalent to the Bowen-York solutions \cite{BowY80} obtained in the context of the CTT decomposition.  This is evident for the choice $\bar \alpha = 1/2$, for which the CTS equations reduce to the corresponding equations in the CTT decomposition and for which the shift $\beta^i$ becomes identical to the Bowen-York vector potential; for other choices of $\bar \alpha$ the shift vector takes a form that results in $\bar u_{ij} = 0$.

\subsection{Wormhole data}
\label{sec:cts_org_worm}

For wormhole data we adopt the Schwarzschild solution on a slice of constant Schwarzschild time, expressed in isotropic coordinates.  The background conformal factor is
\begin{equation}
\psi_0 = 1 + \frac{M}{2r}
\end{equation}
and we have
\begin{equation}
\bar A^{ij}_0 = 0.
\end{equation}
For the background quantities to satisfy equation (\ref{A_1.5}) we therefore have to adopt the Killing shift as our background shift,
\begin{equation}
\beta_0^i = \beta^i_K = 0.
\end{equation}
The Killing lapse of this slice is
\begin{equation} \label{wormhole_Killinglapse}
\alpha_K = \frac{1 - M/(2r)}{1 + M/(2r)}.
\end{equation}
As we discussed in Section \ref{sec:puncture}, this lapse vanishes at the throat at $r = 2M$, so that eq.~(\ref{A_2}) cannot be evaluated there.  Noting that both sides of equation (\ref{A_1.5}) vanish for the background data, we recognize that we can make other choices for the lapse and still have the background quantities satisfy equations (\ref{ham_const_3}) through (\ref{A_1.5}).  Following \cite{HanC05}, we therefore choose the background lapse according to
\begin{equation} \label{lapse_0}
\alpha_0 = \frac{1 + c/(2r)}{1 + M/(2r)},
\end{equation}
where $c$ is a constant.  For $c = -M$ we evidently recover the Killing lapse, but we will choose $c = M$ instead throughout this paper.   Given $\alpha_0$ we compute the densitized background lapse from $\bar \alpha_0 = \alpha_0/\psi_0^6$.  

With our choice of the background lapse (\ref{lapse_0}), the background data satisfy equations (\ref{ham_const_3}) through (\ref{A_1.5}), but they do not satisfy the trace-free part of the time-evolution equation (\ref{K_dot}) with $\partial_t K_{ij} = 0$.  These data would therefore result in a non-trivial time evolution.  It is in this sense that these data do not represent quasiequilibrium data.  To construct quasiequilibrium data we would have to choose the Killing lapse for the background.  While this is not possible for wormhole data, we will see in Section \ref{sec:cts_org_trump} that this is possible for trumpet data.

For the above choice of $\bar \alpha_0$, Laguna \cite{Lag04} found an analytic solution to the momentum constraint (\ref{mom_const_5}) that carries a linear momentum $P^i_L = (0,0,P_L)$ and that, by construction, results in the same extrinsic curvature as the corresponding Bowen-York solution.  Adopting the notation of \cite{HanC05} this solution is given by
\begin{equation} \label{laguna}
\begin{array}{rcl}
\beta^x_L & = & - \displaystyle \frac{x z f(r,M,c)}{5 (M+ 2r)^6} P_L, \\[1mm]
\beta^y_L & = & - \displaystyle \frac{y z f(r,M,c)}{5 (M+ 2r)^6} P_L, \\[1mm]
\beta^z_L & = & - \displaystyle \frac{4 z^2 f(r,M,c) + g(r,M,c)}{20(M+2r)^6} P_L,
\end{array}
\end{equation}
where the functions $f(r,M,c)$ and $g(r,M,c)$ are given by
\begin{eqnarray}
f(r,M,c) & = & (M+c)M^2 + 12 (M+c)Mr  \nonumber \\
&&	+ 60(M+c)r^2 + 160 r^3
\end{eqnarray}
and
\begin{eqnarray}
g(r,M,c) & = & 5(5M+c)M^4 + 60(5M+c)M^3r \nonumber \\
&& 	+ 2(749M + 149c)M^2 r^2 \nonumber \\
&& 	+ 8(497M + 97c)Mr^3 \nonumber \\
&& 	+ 120(49M+9c)r^4 + 4480 r^5
\end{eqnarray}
(see also the red surface in Fig.~\ref{Fig4} below).  This solution serves as a very valuable test for our numerical code.

In our finite-difference code, we use Cactus and PETSc software to invert the operator on the left-hand side of eq.~(\ref{mom_const_5}) simultaneously for all three components of $b^i$.  We adopt Cartesian coordinates with $N$ equidistant gridpoints in each spatial dimension.  At the outer boundaries, imposed at identical locations $X_{\rm out} = Y_{\rm out} = Z_{\rm out}$, we enforce the fall-off conditions of the Laguna solution (\ref{laguna}).   Wormhole data satisfy certain symmetry conditions, so that we could restrict our code to one octant only and impose symmetry conditions on the coordinate planes.  However, for the trumpet data in Section \ref{sec:cts_org_trump} these conditions to not apply; we therefore will not use these symmetry conditions here either.

Since eq.~(\ref{mom_const_5}) is linear, solutions cannot be unique.  We pick a particular solution by fixing the components of $b^i$ at the puncture $r=0$, where eq.~(\ref{mom_const_5}) becomes singular
\begin{equation} \label{fix_point}
b^i_{\rm punc} = -\frac{5M+c}{4M^2}P^i_L
\end{equation}
(see \cite{HanC05} for a more detailed discussion and motivation for this approach).  We choose a vertex-centered grid, so that the puncture coincides with a grid point, and we force the solution to take the value (\ref{fix_point}) there.  

\begin{figure}[t]
\begin{center}
\includegraphics[width=3in]{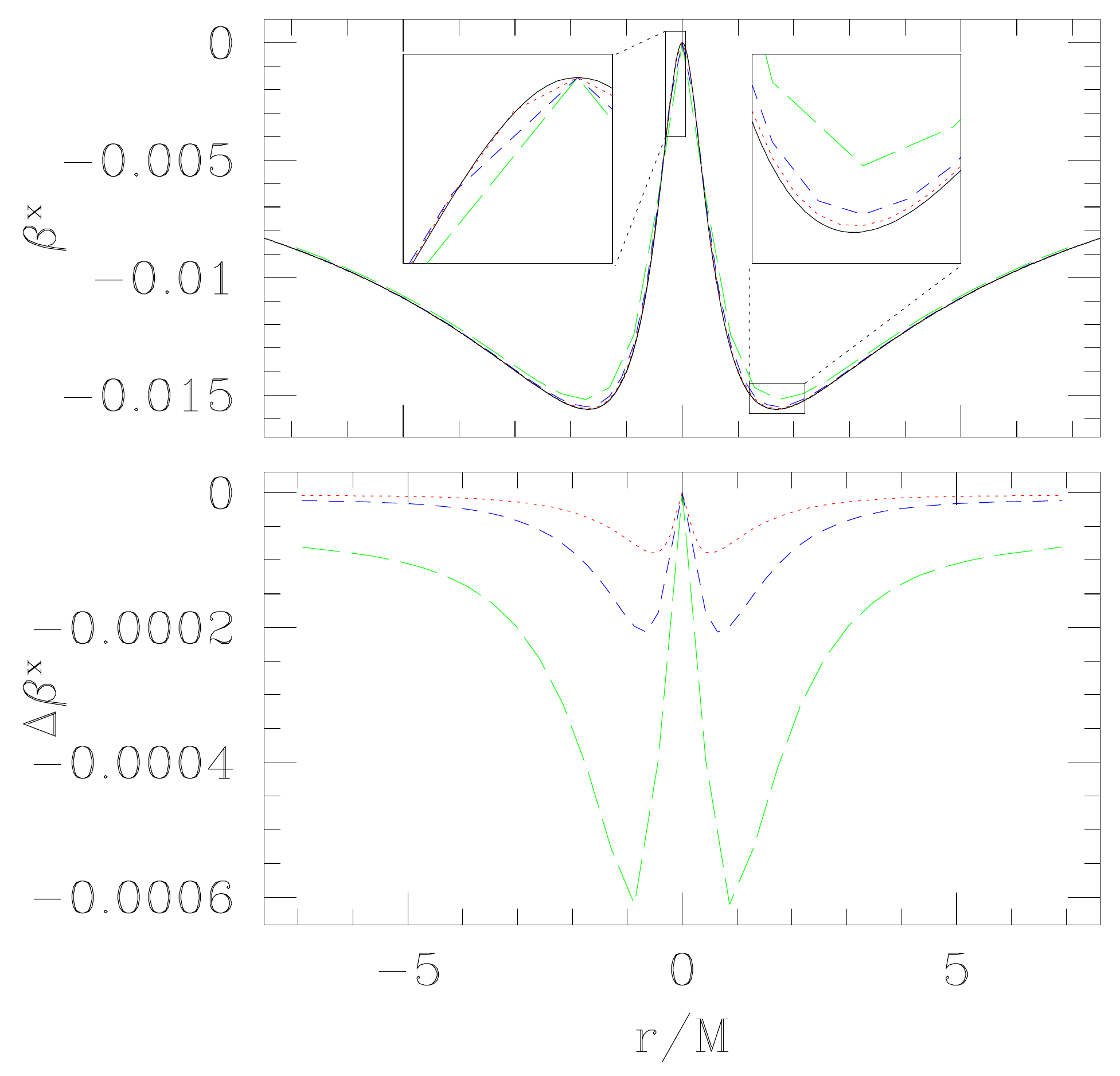}
\end{center}
\caption{Convergence of the shift component $\beta^x$ to the Laguna solution (\ref{laguna}) for a boost of $P = 0.5M$ and $c = M$.  Here the outer boundaries were imposed at $X_{\rm out} = 4M$, and we show results along the space diagonal for three resolutions.  The upper panel shows the convergence of the numerical solutions to the analytical result, shown as the solid line, and the lower panel shows the differences between the analytical solution and the numerical results.  The insets show enlargements of the regions marked by the boxes.}
\label{Fig2}
\end{figure}

As an example, we show some numerical results for the $x$ component of the shift in Fig.~\ref{Fig2}, demonstrating that our numerical solutions converge to the analytical solution.  In agreement with the findings of \cite{HanC05}, the rate of convergence is between first and second order; the lack of second-order convergence is caused by the singular nature of the puncture, where we force the solution to take
fixed values.

\begin{figure}[t]
\begin{center}
\includegraphics[width=3in]{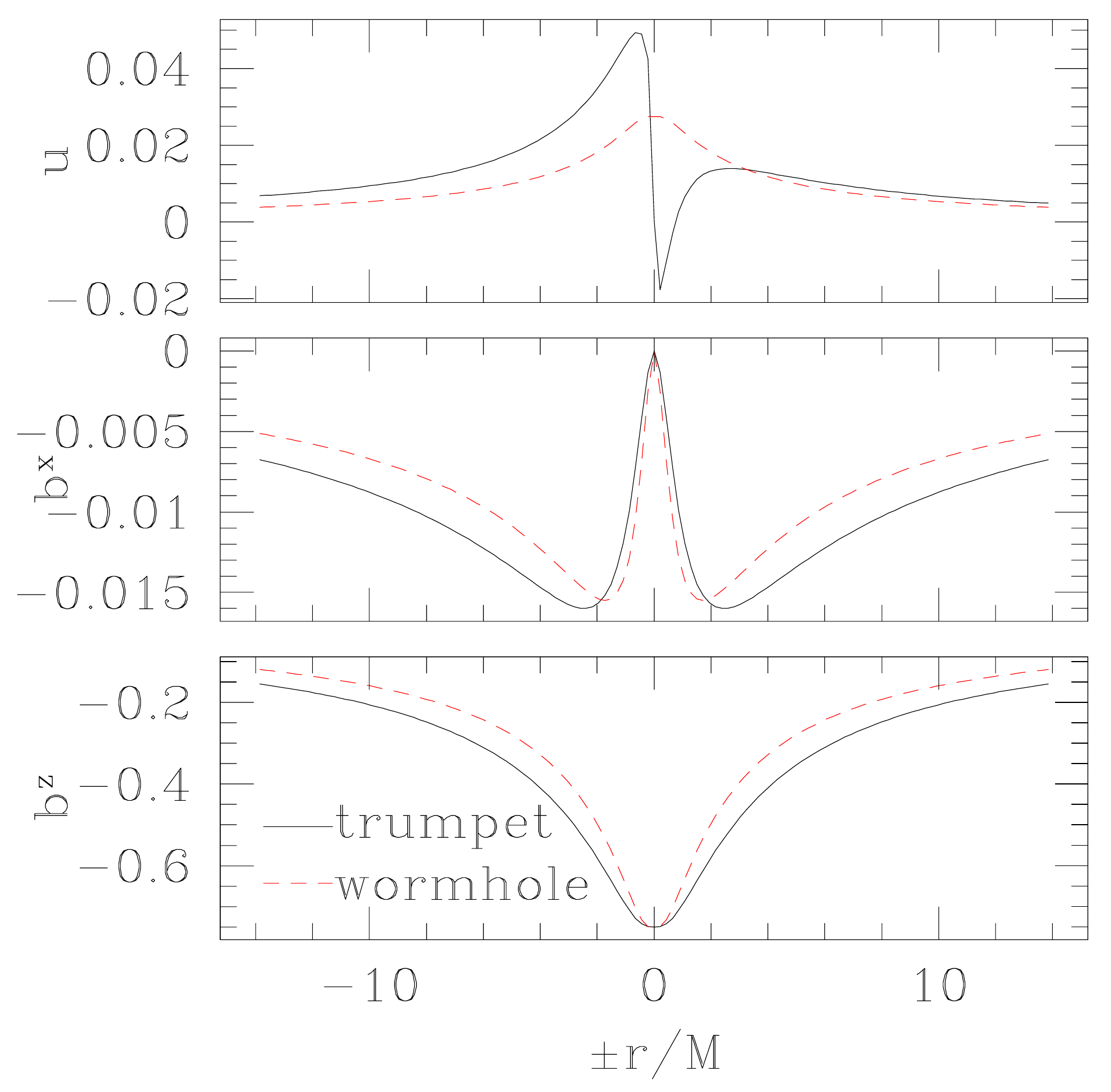}
\end{center}
\caption{Numerical results for the corrections $u$, $b^x$ and $b^z$, for both wormhole and trumpet data in the original CTS formalism, for $P_L = 0.5 M$.  Results are shown along the space diagonal, with outer boundaries imposed at $X_{\rm out} = 8M$ and $N=129$ gridpoints in each dimension.}
\label{Fig3}
\end{figure}

Given a solution for $b^i$, we can compute $\bar A^{ij}$ from (\ref{A_2}) and insert the result into the Hamiltonian constraint (\ref{ham_const_4}).  For wormhole data, the right-hand side approaches zero at the puncture, so that the equation can be solved without special treatment of the puncture.   The Hamiltonian constraint is non-linear in $u$ and cannot be solved directly.  We linearize the equation in corrections to $u$ and solve the resulting equation iteratively until the L2 norm of the residual has dropped below a certain tolerance (here chosen to be $10^{-9}$ in code units).   We show results for a boost of $P = 0.5M$ in Fig.~\ref{Fig3}.

\subsection{Trumpet data}
\label{sec:cts_org_trump}

For trumpet data we adopt the Schwarzschild solution on a trumpet slice as background data.  More specifically, we adopt the limiting case of the family of maximal slices that we described in Section \ref{sec:puncture}.  Maximally sliced trumpet data are discussed in \cite{HanHOBGS06}, and an analytic solution in isotropic coordinates is given in \cite{BauN07}.  We summarize this solution in \ref{AppA}.

The background conformal factor $\psi_0$ and the background extrinsic curvature $\bar A^{ij}$ are given by (\ref{psi_trump}) and (\ref{A_trump}).  As before we adopt the Killing shift, now given by (\ref{shift_trump}), as the background shift $\beta^i_0$.  Since neither $\bar A^{ij}_0$ nor $\beta^i_0$ vanish everywhere, we now have to adopt the Killing lapse (\ref{alpha_trump}) as the background lapse for the background quantities to satisfy eq.~(\ref{A_1.5}).  Unlike for wormhole data, this does not pose a problem, since this Killing lapse remains positive for all $r >0$ (see also Fig.~2 in \cite{BauN07}).    Adopting the Killing lapse as the background lapse will lead to quasiequilibrium data, which represents a potential advantage over the approach for wormhole data in Section \ref{sec:cts_org_worm}.   As we discussed in Section \ref{sec:puncture}, this observation motivates the study of trumpet-puncture initial data in the CTS formalism.

We can now solve equations (\ref{ham_const_4}) - (\ref{mom_const_5}) with trumpet background data.  Unlike for wormhole data, the right hand side of the Hamiltonian constraint (\ref{ham_const_4}) does not remain finite at the puncture for a trumpet background.   For non-spinning black holes this implies that the correction $u$ has to vanish at the puncture itself (see \cite{ImmB09,HanHO09}).  Instead of solving the Hamiltonian constraint everywhere, we therefore fix the solution to zero at the puncture, similar to how we force the shift to take certain values at the puncture.

\begin{figure}[t]
\begin{center}
\includegraphics[width=3in]{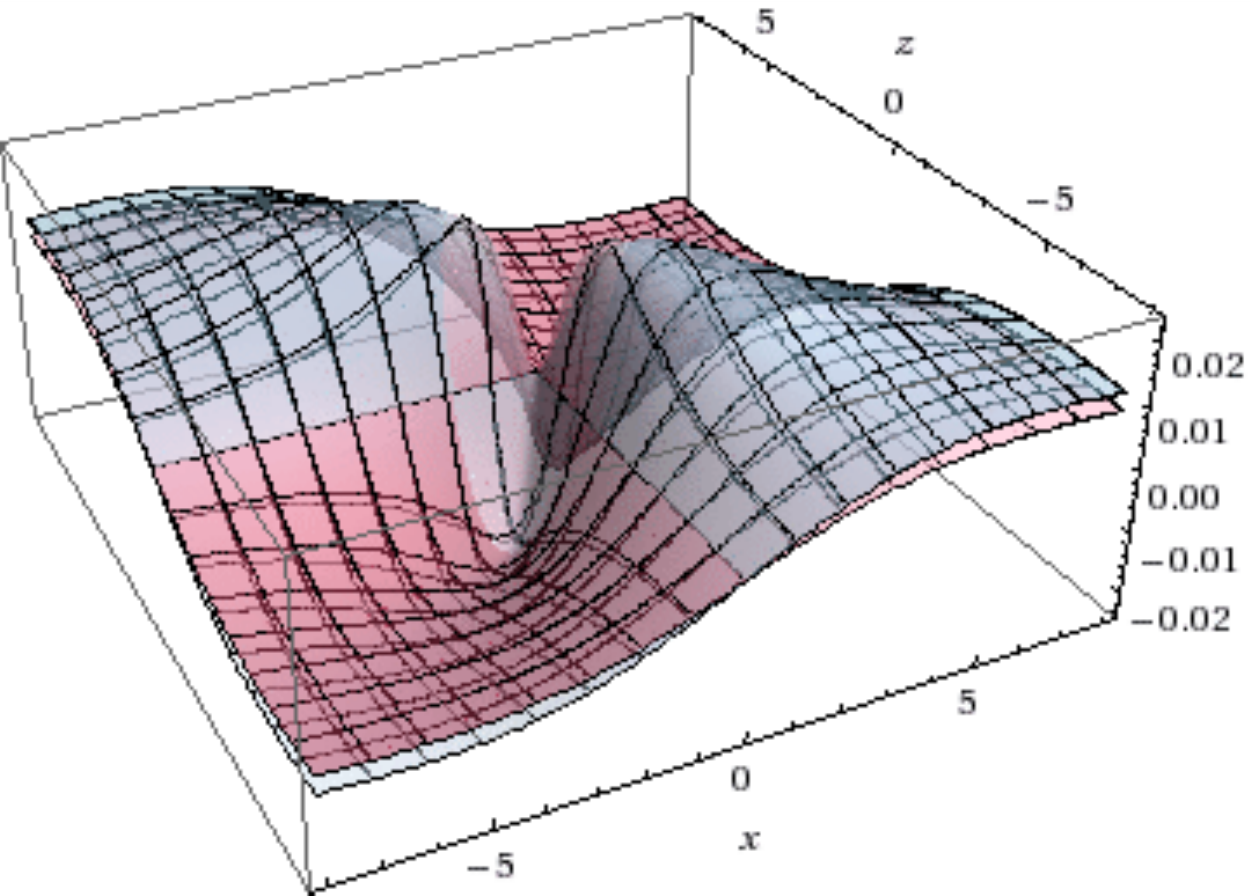}
\end{center}
\caption{A comparison of the $b^x$-component for wormhole and trumpet data, for a boost of $P_L = 0.5M$.  The blue data (on top on the right and left) represent trumpet data, while the red data (on top in the front and back) represent wormhole data.}   
\label{Fig4}
\end{figure}

In Fig.~\ref{Fig3} we show results for the corrections to the conformal factor and the shift, and compare with the corresponding results for wormhole data.  To specify the solution for the shift, we again set the shift correction $b^i$ to (\ref{fix_point}) at the puncture - the wormhole and trumpet data therefore agree at the puncture.   Since the Laguna solution (\ref{laguna}) is not a solution to equation (\ref{mom_const_5}) with trumpet background data, the two solutions are not identical away from the puncture.  However, as shown in Fig.~\ref{Fig3}, the difference between the two solutions is not dramatic.  This is also evident in the surface plot in Fig.~\ref{Fig4}, where we show the $x$ component of $b^i$, for both wormhole and puncture data, in the $x$-$z$ plane.

For trumpet data, the linear momentum is no longer given by the parameter $P_L^i$ in (\ref{fix_point}).  We instead compute the linear momentum from
\begin{equation} \label{lin_mom}
P^i = \frac{1}{8 \pi} \oint \bar A^{ij} d^2 S_j,
\end{equation}
and evaluate the integral on the outer boundary of our computational domain.  Given that the momentum constraint (\ref{mom_const_5}) is linear, we now find the linear relation
\begin{equation}
P^i = 1.43 P^i_L
\end{equation}
for our choice of $c = M$.  
 
\section{Discussion}
\label{sec:discussion}

We reconsider the construction of quasiequilibrium puncture black holes in the context of the CTS decomposition.  It has been shown previously that this approach cannot be used for wormhole data, which connect the two spatial infinities in two separate universes (see \cite{HanECB03}).  For such data, the Killing lapse has to vanish somewhere away from the puncture even for Schwarzschild black holes, so that the CTS formalism cannot be applied (this problem can be avoided by choosing a different lapse function, which, however, does not lead to quasiequilibrium data, see \cite{HanC05,Han05}).  Here we point out that this argument does not apply to trumpet data, which connect the spatial infinity in one universe with the future timelike infinity of another.   For these data, the Killing lapse vanishes only at the puncture itself, which suggests that it may be possible to construct quasiequilibrium trumpet-puncture data in the CTS formalism.

We verify that it is possible to construct such quasiequilibrium trumpet-puncture data in the original CTS formalism, and present numerical results for single boosted black holes.   We also compare these trumpet data with wormhole data that have been constructed using the same formalism, but with a non-Killing lapse.

We also attempted to solve the equations of the extended CTS formalism for both wormhole and puncture data.  The equations appear similar to those of the original CTS formalism, except that we now solve the lapse equation (\ref{lapse_1}) together with the other equations.  This, however, introduces an important difference; instead of fixing $\bar \alpha$, as in the original CTS formalism, we now fix $\partial_t K$ and compute $\bar \alpha$ from $\psi$ and $\phi$.  Inserting this dependency explicitly in equations (\ref{ham_const_2}) and (\ref{lapse_1}) changes the sign of the exponents of $\psi$ and $\phi$ in some of the terms.  As a consequence, the maximum principle can no longer be applied to establish uniqueness of solutions (see \cite{BauMP07,Wal07}).

We have solved the equations of the extended CTS formalism, for a wormhole background with a non-Killing lapse, to reproduce the results of \cite{HanC05}.   These solutions exist only up to a certain critical value of the momentum (compare Fig.~8 in \cite{HanC05}), and may not be unique (compare \cite{PfeY05,BauMP07}).  For trumpet data, however, we have been unable to find any converging solutions, even for very small values of the momentum.   While we cannot rule out that the fault lies with our numerical code, we suspect that these solutions do not exist.  The equations form a complicated system of non-linear equations, especially for a trumpet background, and we will postpone a further investigation of whether these equations allow any regular non-trivial solutions  to a future study.

However, even the trumpet data constructed within the framework of the original CTS formalism are an interesting alternative to more traditional black hole initial data; as trumpet data they are closer to the geometry realized in dynamical moving-puncture simulations than wormhole data, and as CTS data they are represented in a coordinate system that is well suited for numerical evolution.

\ack

I would like to thank Mark Hannam for useful conversations, and Jason Immerman for his contributions to the early part of this project.  This work was supported in part by NSF grant
PHY-0756514 to Bowdoin College.

\begin{appendix}

\section{The Schwarzschild solution in a maximally sliced trumpet geometry}
\label{AppA}

A family of maximal slices of the Schwarzschild spacetime can be given analytically in terms of the areal radius $R$ (see \cite{EstWCDST73,Rei73,BeiO98}).  The limiting member of this family, which describes a trumpet geometry, can also be transformed analytically into isotropic coordinates, even though the solution can only be given parametrically, with $R$ as the parameter \cite{BauN07}.  In our numerical implementation, we determine all variables as a function of $r$ by creating tables and using suitable interpolation.

In particular, the conformal factor is given by
\begin{eqnarray} \label{psi_trump}
\psi
&=&
\left[ 
\frac
{4 R} 
{2 R +  M + (4 R^2 + 4 M R + 3 M^2)^{1/2} }
\right]^{1/2} \\
&  & \times 
\left[ 
\frac
{8 R +  6 M + 3 ( 8 R^2 + 8 M R + 6 M^2 )^{1/2} } 
{(4 + 3 \sqrt{2})(2 R - 3 M) }
\right]^{1/2\sqrt{2}} 
\nonumber
\end{eqnarray}
where the isotropic radius $r$ is
\begin{eqnarray} \label{eq:r}
r & = & \left[ 
\frac{2 R +  M + (4 R^2 + 4 M R + 3 M^2)^{1/2} } {4} 
\right]  \\
& &  \times 
\left[ \frac{(4 + 3 \sqrt{2})(2 R - 3 M) }
{8 R +  6 M + 3 ( 8 R^2 + 8 M R + 6 M^2 )^{1/2} } 
\right]^{1/\sqrt{2}} . \nonumber
\end{eqnarray}
Asymptotically, $\psi_0$ behaves as
\begin{equation}
\psi = \left\{ \begin{array}{ll}
\displaystyle \left( \frac{3M}{2r} \right)^{1/2} ~~~~~ & r \rightarrow 0, \\[3mm]
\displaystyle 1 + \frac{M}{2r} &  r \rightarrow \infty.
\end{array} \right.
\end{equation}
The limit surface $r \rightarrow 0$ corresponds to a sphere of areal radius $R = \psi^2 r \rightarrow 3 M /2$.  

The Killing lapse and shift are given by
\begin{equation} \label{alpha_trump}
\alpha_K = \left(1 - \frac{2M}{R} + \frac{27M^4}{16R^4} \right)^{1/2}
\end{equation}
and
\begin{equation} \label{shift_trump}
\beta^r_K = \frac{3\sqrt{3}M^2}{4}\, \frac{r}{R^3}.
\end{equation}
To determine the behavior of $\alpha_K$ in the neighborhood of the puncture, we write
\begin{equation}
R = \frac{3M}{2} + \rho
\end{equation}
and expand both (\ref{eq:r}) and (\ref{alpha_trump}) to leading order in $\rho$.   This results in
\begin{equation} \label{r_exp}
r \simeq \kappa M \left( \frac{2 \kappa}{9} \frac{\rho}{M} \right)^{1/\sqrt{2}},
\end{equation}
where we have abbreviated $\kappa = (4 + 3 \sqrt{2})/4$, and
\begin{equation} \label{alpha_exp}
\alpha_K \simeq \frac{2\sqrt{2}}{3} \frac{\rho}{M}.
\end{equation}
Combining eqs.~(\ref{r_exp}) and (\ref{alpha_exp}) we can eliminate $\rho$ and find that, to leading order, $\alpha_K$ scales with $r^{\sqrt{2}}$,
\begin{equation} \label{alpha_exp_2}
\alpha_K \simeq \frac{3 \sqrt{2}}{\kappa} \left(\frac{1}{\kappa} \frac{r}{M} \right)^{\sqrt{2}}.
\end{equation}

Finally, the conformally related, trace-free part of the extrinsic curvature is
\begin{equation} \label{A_trump}
\bar A^{ij} = \frac{3 \sqrt{3} M^2}{4 r^3}(\bar \gamma^{ij} - 3 n^i n^j),
\end{equation} 
where $n^i = x^i/r$ is the spatial normal vector pointing away from the puncture at $r=0$.

\end{appendix}

     
\end{document}